\documentclass[fleqn,usenatbib]{mnras}

\DeclareRobustCommand{\VAN}[3]{#2}
\let\VANthebibliography\thebibliography
\def\thebibliography{\DeclareRobustCommand{\VAN}[3]{##3}\VANthebibliography}

\usepackage{graphicx}	
\usepackage{amsmath}	
\usepackage{amssymb}	

\title[Galaxy Spin Asymmetry]{Analysis of spin directions of galaxies in the DESI Legacy Survey}

\author[Lior Shamir]{
Lior Shamir,$^{1}$\thanks{E-mail: lshamir@mtu.edu}
\\
$^{1}$Kansas State University, Manhattan, KS, 66506, USA\\
}

\date{Accepted 2022 August 18. Received 2022 July 15; in original form 2022 April 28}

\pubyear{2022}

\begin{document}
\label{firstpage}
\pagerange{\pageref{firstpage}--\pageref{lastpage}}
\maketitle

\begin{abstract}
The DESI Legacy Survey is a digital sky survey with a large footprint compared to other Earth-based surveys, covering both the Northern and Southern hemispheres. This paper shows the distribution of the spin directions of spiral galaxies imaged by DESI Legacy Survey. A simple analysis of dividing nearly 1.3$\cdot10^6$ spiral galaxies into two hemispheres shows a higher number of galaxies spinning counterclockwise in the Northern hemisphere, and a higher number of galaxies spinning clockwise in the Southern hemisphere. That distribution is consistent with previous observations, but uses a far larger number of galaxies and a larger footprint. The larger footprint allows a comprehensive analysis without the need to fit the distribution into an a priori model, making this study different from all previous analyses of this kind. Fitting the spin directions of the galaxies to cosine dependence shows a dipole axis alignment with probability of $P<10^{-5}$. The analysis is done with a trivial selection of the galaxies, as well as simple explainable annotation algorithm that does not make use of any form of machine learning, deep learning, or pattern recognition. While further work will be required, these results are aligned with previous studies suggesting the possibility of a large-scale alignment of galaxy angular momentum.
\end{abstract}




\section{Introduction}
\label{introduction}

According to the {\it tidal torque theory}, the angular momentum of galaxies is initiated by subtle misalignment between the tidal shear tensor and the Lagrangian patch during a period of gravitational instability \citep{hoyle1949problems,peebles1969origin,doroshkevich1970spatial,efstathiou1979rotation,white1984angular,catelan1996evolution,pen2000tentative,lee2001galaxy,porciani2002testing,porciani2002testing2,schafer2009galactic}. Tidal torque theory was first proposed by \cite{hoyle1949problems}, but according to the literature was largely ignored by the scientific community until echoed by \cite{peebles1969origin} about two decades later \citep{efstathiou2005fred}. The theory explains the initial origin of angular momentum of galaxies, making a link between the large-scale structure and the alignments of galaxy spin, as both are related to the same large-scale tidal force field that includes the gravitational collapse that determined the structure of the filaments and walls that make the cosmic web \citep{lee2000cosmic,porciani2002testing,porciani2002testing2,schafer2009galactic}.

While early observations found no statistically significant alignment of spin directions in cosmic filaments \citep{han1995orientation}, the deployment of powerful autonomous digital sky surveys has enabled the collection of far larger astronomical databases that were not available the pre-information era of astronomy research. The presence and accessibility of these databases provided new tools for the studying of the large-scale structure. By using data collected by the Sloan Digital Sky Survey and the Two-Micron All-Sky Survey, a link between the alignment of galaxy spin directions and the large-scale structure was identified \citep{jones2010fossil}. That study was followed by numerous observations, showing a large-scale alignment in the spin directions of galaxies \citep{jones2010fossil,tempel2013evidence,tempel2013galaxy,tempel2014detecting,pahwa2016alignment,ganeshaiah2018cosmic,lee2018wobbling,ganeshaiah2019cosmic,lee2019galaxy,lee2019mysterious,blue2020chiles,welker2020sami,kraljic2021sdss,lopez2021deviations,motloch2021observed}.  

These empirical observations are also aligned with numerical simulations, showing a link between the large-scale structure and the distribution of spin directions of galaxies \citep{zhang2009spin,davis2009angular,libeskind2013velocity,libeskind2014universal,forero2014cosmic,wang2018build,lopez2019deviations}. A direct statistical correlation was also found between the spin directions of galaxies and cosmic initial conditions \citep{motloch2021observed}. Despite these advancements and substantial research efforts, the alignments of galaxy rotations are still not fully understood \citep{lopez2021deviations}.

Another explanation to galaxy rotation is that angular momentum of galaxies is driven by galaxy mergers \citep{vitvitska2002origin,peirani2004angular}. Unlike tidal torque theory, the merging of galaxies is expected to lead to more stochastic spin of galaxies that is not directly related to the initial conditions \citep{cadiou2021angular}. It has been also suggested that dark matter halos passing close to each other can also initiate rotation \citep{ebrahimian2021dynamical}. On the other hand, certain observations suggest that rotating star-forming disk galaxies existed in the early Universe, and before they could be formed through gravitational interactions \citep{neeleman2017c,neeleman2020cold}. 

\cite{welker2020sami} made use of galaxies from the Sydney-Australian-Astronomical-Observatory Multi-object Integral-Field Spectrograph (SAMI) survey. By using a dataset of 1,418 galaxies, the study showed a statistically significant spin alignment with cosmic filaments, and dependence on stellar mass. The results showed evidence of initial spin of galaxies of lower mass aligned with the parent structure, followed by mergers that flip the spin direction, leading to perpendicular alignment in comparison to the parent structure. The dependency on mass was first observed through numerical simulations \citep{aragon2007spin}, with numerous consequent studies that showed similar results \citep{hahn2007properties,codis2012connecting,trowland2013spinning,casuso2015origin,wang2017general,wang2018spin,ganeshaiah2018cosmic,ganeshaiah2019cosmic,ganeshaiah2021cosmic}.



First observations that suggested that the distribution of spin directions of spiral galaxies was not necessarily random were based on manual annotation of the galaxies \citep{macgillivray1985anisotropy,longo2011detection,lee2019mysterious}. The ability to annotate galaxy images automatically by using machine vision to determine their spin directions allowed the analysis of a far larger number of galaxies \citep{shamir2012handedness}. By using a fully symmetric algorithm, computer analysis has the advantage of not being subjected to the effect of the human perceptual bias. Using a large number of automatically annotated galaxies covering the Northern \citep{shamir2020patterns} and Southern \citep{shamir2021large} hemispheres also allowed to profile the nature of the non-random patterns \citep{shamir2022new,shamir2022large}. These patterns are consistent across different telescopes \citep{shamir2022possible} such as SDSS \citep{shamir2019large,shamir2020patterns}, HST \citep{shamir2020pasa}, Pan-STARRS \citep{shamir2020patterns} and DECam \citep{shamir2021large}, showing very similar profiles \citep{shamir2022large}. The patterns are also similar whether using automatic annotation or manual annotation of the galaxies \citep{shamir2020pasa}.

On the other hand, other studies argued that the spin direction of spiral galaxies are distributed randomly. \cite{iye1991catalog} prepared a catalogue of over $6\cdot10^3$ and showed that the distribution of spin directions of the galaxies was random. However, when the asymmetry is $<$1\%, a set of $6.5\cdot10^3$ galaxies is far too small to provide statistically significant non-random distribution of such mild magnitude. Using binomial distribution, at least $2.7\cdot10^4$ galaxies are required to show a P value of less than 0.05 when the asymmetry in the data is 1\%.

An attempt to use non-expert volunteers to annotate the galaxies by their spin direction showed a random distribution \citep{land2008galaxy}. The experiment also showed that the human annotators had a very strong bias, driven by the human perception \citep{land2008galaxy}. When the bias was noticed, another experiment was done by mirroring a relatively small subset of the images to correct for the human bias. As shown in Table 2 in \citep{land2008galaxy}, when the images were mirrored the number of counterclockwise galaxies was reduced from 6.032\% counterclockwise galaxies to 5.942\% mirrored clockwise galaxies, showing a $\sim$2\% higher number of counterclockwise galaxies. At the same time, the number of clockwise galaxies increased from 5.525\% clockwise galaxies to 5.646\% mirrored counterclockwise galaxies, again showing a 1.5\% more counterclockwise galaxies. The small size of the dataset does not provide statistically significant results (P$\simeq$0.13), but the difference of 1.5\%-2\% fewer galaxies that spin clockwise in the footprint of SDSS galaxies with spectra agrees with the results of automatically annotated galaxies from the exact same footprint \citep{shamir2020patterns}.

Another analysis that determined that the distribution was random used a computer algorithm to annotate the images \citep{hayes2017nature}. The use of computer vision allowed to annotate a larger number of galaxies, and also avoid the strong human perceptual bias that can impact the process and is difficult to correct. As specified in Table 2 in \citep{hayes2017nature}, the results of the analysis showed asymmetry between the number of galaxies spinning in opposite directions, with probability of 2.52$\sigma$ to occur by chance in the population of SDSS galaxies that were annotated manually as spirals. To avoid a possible bias in the galaxy selection, a machine learning algorithm was applied for the selection of the spiral galaxies, and the machine learning algorithm also provided a dataset with a higher number of galaxies spinning counterclockwise. To make the distribution random, the machine learning algorithm was modified such that the attributes that correlate with the galaxy spin direction asymmetry were removed. That is explained in Section 4.1 of \citep{hayes2017nature}. After selecting the galaxies by removing specifically the attributes that correlate with the asymmetry in galaxy spin direction, the asymmetry in the galaxies that were selected by the machine learning algorithm became lower, regardless of whether the source of the asymmetry was the sky or a certain bias in the algorithm.

\cite{iye2020spin} proposed that the possible asymmetry between galaxies with opposite spin directions is the result of duplicate objects they believed existed in the dataset \citep{iye2020spin}. They showed that after removing ``duplicate objects" the statistical significance of the dipole axis dropped to 0.29$\sigma$. However, the dataset they used for that purpose \citep{shamir2017photometric} was an older dataset of photometric objects that was designed and used for photometric analysis \citep{shamir2017photometric}. No claim for the presence or absence of any kind of axis was made in \citep{shamir2017photometric}, and no such claim about that dataset was made in any other paper. When using that dataset to count the number of galaxies, photometric objects that are part of the same galaxy such as merging galaxies, large star clusters, satellite galaxies, etc become duplicate objects. But as mentioned above, no claim for any kind of axis in that dataset was made in \citep{shamir2017photometric} or in any other paper.

But in addition to removing duplicate objects, the 0.29$\sigma$ signal was observed after also limiting the galaxies to $z_{phot}<0.1$. That observation of random distribution in $z<0.1$ is in full agreement with previous work \citep{shamir2020patterns}, showing that the asymmetry signal becomes stronger as the redshift gets higher. Namely, Tables 3, 5, 6 and 7 in \citep{shamir2020patterns} show random distribution at $z<0.1$, and therefore the results of \citep{iye2020spin} are in agreement with previous reports \citep{shamir2020patterns}. 

When not limiting to lower redshifts, the dataset used in \citep{iye2020spin} shows non-random distribution of the spin directions of the galaxies, and a statistically significant dipole axis. The exact same dataset used by \cite{iye2020spin} is available at \url{https://people.cs.ksu.edu/~lshamir/data/assym_72k/}. As shown in Table~\ref{hemispheres}, a very simple analysis using simple binomial distribution shows that the dataset can be separated into two hemispheres, such that one hemisphere has a higher number of galaxies spinning clockwise, and the opposite hemisphere has a higher number of galaxies spinning counterclockwise. Even when assuming that the distribution in the less populated hemisphere is fully random, a Bonferroni correction still provides probability of $\sim0.01$ of the asymmetry to occur by chance. 

\begin{table}
\caption{The distribution of galaxies spinning clockwise and counterclockwise in the exact same dataset used in \citep{iye2020spin}. The P values are the one-tailed and two-tailed P values based on binomial distribution when assuming that the probability of a galaxy to spin either clockwise or counterclockwise is mere chance 0.5.}
\label{hemispheres}
\scriptsize
\begin{tabular}{lcccc}
\hline
Hemisphere & \# cw          & \# ccw         &  $\frac{\#Z}{\#S}$  & P  \\ 
   (RA)         &                        &                         &                     & (one-tailed)  \\ 
\hline
$70^o-250^o$               & 23,037 & 22,442   &   1.0265   &  0.0026 \\ 
$>250^o \cup <70^o$   &  13,660 &  13,749  &   0.9935   &  0.29   \\ 
\hline
\end{tabular}
\end{table}

Repeating the exact same analysis to identify a dipole axis as described in Equation 2 in \citep{iye2020spin} with the exact same data that was used in that experiment but without limiting the redshift shows a dipole axis with statistical significance of 2.14$\sigma$, and therefore can be considered statistically significant. The exact same data used in \citep{iye2020spin}, the code implementing the analysis, instructions of running the code, and the resulting output of the analysis are available in \url{https://people.cs.ksu.edu/~lshamir/data/iye_et_al}.

Here, a large dataset of nearly 1.3$\cdot10^6$ galaxies annotated by their spin direction is studied. The dataset is based on galaxy images acquired by the DESI Legacy Survey, which is unique in the size of its footprint, and its ability to cover both the Northern and the Southern hemisphere.

\section{Data}
\label{data}

The DESI Legacy Survey \citep{dey2019overview} is a digital sky survey that makes use of three different instruments. These include the Dark Energy Camera (DECam), the Beijing-Arizona Sky Survey (BASS), and the Mayall z-band Legacy Survey (MzLS). Through image calibration, the sky survey provides a dataset of almost uniform depth \citep{dey2019overview}. The combination of three different instruments located in both the Nortehrn and Southern hemispheres provides a footprint of over 14,000 degrees$^2$.

The list of objects was based on DESI Legacy Survey Data Release (DR) 8. Objects with g magnitude brighter than 19.5 and identified by the DESI Legacy Survey photometric pipeline as round exponential galaxies (`'REX''), exponential disks (``EXP''), or de Vaucouleurs ${r}^{1/4}$ profiles (``DEV''), were selected as the initial set of objects. The purpose of the selection was the reduce the DESI Legacy Survey objects to a shorter list of objects identified as extended sources, but also to exclude objects that are embedded in other extended sources.

The images were retrieved automatically by using the {\it cutout} API service of the DESI Legacy Survey server. The images were of dimensionality of 256$\times$256, and in the JPEG format. To ensure the entire object fits in the frame, the images were scaled by the Petrosian radius. The total number of images that were retrieved was 34,475,215. Retrieving the image data from declination range $\delta<30^o$ started on June 4th, 2020, and continued until March 4th, 2021. Retrieving the galaxy images at declination higher than 30$^o$ started on November 15th 2021, and ended on March 7th 2022. The process of image data retrieval required more than a year of continuous downloading of data from the DESI Legacy Survey server.


\subsection{Annotation of the galaxies by their spin direction}
\label{annotation}

As discussed in Section~\ref{introduction}, computer annotation of the galaxies allows to annotate large datasets, while also avoid the bias of the human perception. Pattern recognition, machine learning, and in particular deep learning rely on manually annotated images from which the machine learning algorithm can be trained, and even the order by which the training samples are used can lead to different rules. The training process leads to complex non-intuitive rules that are very difficult to analyse or to prove their symmetry. That can lead to biases of these algorithms that are difficult to control or profile \citep{dhar2022systematic}. Empirical experiments can very often be misleading \citep{carter2020overinterpretation,dhar2021evaluation}, and can specifically affect the annotation of galaxy images \citep{dhar2022systematic}. Due to the bias and the difficulty to control the consistency of the annotation, machine learning should not be considered a valid approach for a study of this kind.

To ensure that the analysis is symmetric, the model-driven Ganalyzer algorithm was used \citep{shamir2011ganalyzer}. {\it Ganalyzer} is not based on any aspect of machine learning, deep learning, or pattern recognition. It is a model-driven deterministic algorithm that follows clear and symmetric rules. While model-driven algorithms can also be biased, the absence of complex data-driven rules makes these algorithms easier to understand and analyse. The use of clear and defined rules therefore allows to provide analysis of the symmetry of the algorithm. Such algorithms can be biased, but the use of defined rules reduces the possibility of asymmetry in the algorithm. Theoretical and empirical analysis of the symmetric nature of the algorithm is provided in \citep{shamir2011ganalyzer,shamir2020patterns,shamir2020pasa,shamir2021large,shamir2022new,shamir2022large}.  

In summary, the Ganalyzer algorithm works by first applying a median filter with a window size of 5$\times$5, and then converting each galaxy image into its radial intensity plot transformation. The radial intensity transformation is defined by Equation~\ref{radial_intensity_plot_equation}

\begin{equation}
\label{radial_intensity_plot_equation}
I_{x,y}=I_{O_x+\sin(\theta) \cdot r ,O_y-\cos(\theta)\cdot r}, 
\end{equation}

where $I_{x,y}$ is the intensity of pixel $(x,y)$ in the radial intensity plot transformation, $O_x$, $O_y$ are the pixel coordinates of the centre of the galaxy in the original galaxy image, $\theta$ is the polar angle of the pixel compared to ($O_x$, $O_y$), and {\it r} is the radial distance.

\begin{figure}
\centering
\includegraphics[scale=0.40]{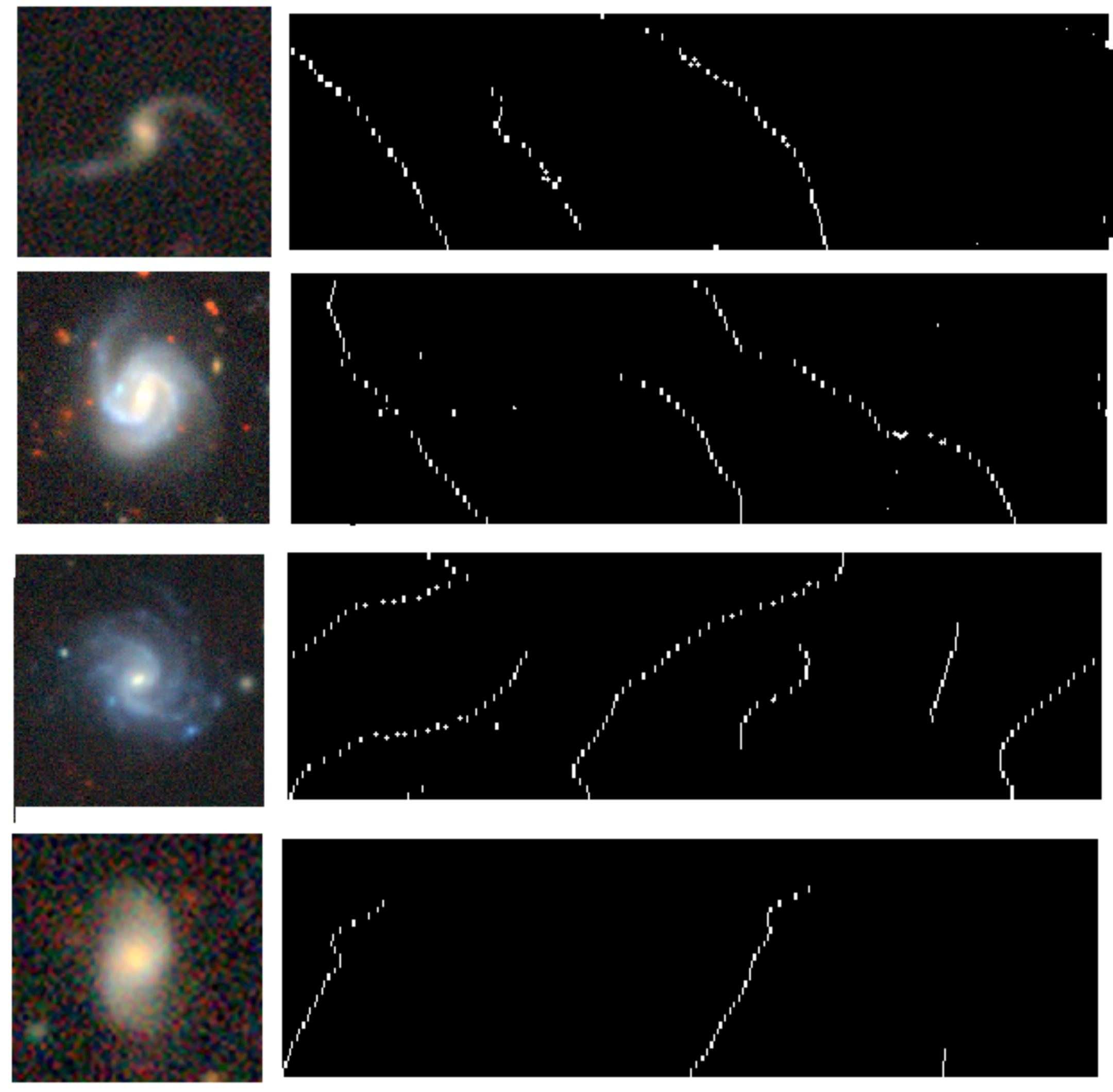}
\caption{Examples of the peaks of the radial intensity plots of different galaxy images from the DESI Legacy Survey. The direction of shift in the location of the peaks corresponds to the curves of the galaxy arms, and therefore can be used to determine the direction of the curve. }
\label{radial_intensity_plot}
\end{figure}

After converting the galaxy image into its radial intensity plot transformation, a peak detection algorithm \citep{morhavc2000identification} is applied to all horizontal lines to detect peaks in each line. The galaxy arms are brighter than other pixels at the same distance from the galaxy centre, and therefore the peaks in the radial intensity plot correspond to the galaxy arms in the original galaxy image. When the arms of a spiral galaxy are curved, the peaks are expected to form a line towards the direction of the curve of the arm. Figure~\ref{radial_intensity_plot} shows several examples of galaxy images (left) and the peaks of the radial intensity plot transformation (right) of each galaxy.

When applying a linear regression to each line, the sign of the regression determines the direction towards which the arm is curved, and consequently the direction towards which the galaxy spins. The linear regression slope $\beta$ formed by these points is determined simply by the value of $\beta$ that satisfies $\min \Sigma_i (r_i- \beta \cdot \theta_i + \epsilon)^2$. If the slope $\beta$ is positive, the galaxy can be determined as spinning clockwise, while if $\beta$ is negative, the galaxy is a counterclockwise galaxy. The Ganalyzer method is explained in detail including experimental results in \citep{shamir2011ganalyzer,shamir2017photometric,shamir2020patterns,shamir2021large,shamir2022new,shamir2022large}. The importance of the symmetricity of the algorithm and the analysis of bias is described in \citep{shamir2021particles}, and briefly in Section~\ref{symmetry}.

Since the galaxy images in the initial set were retrieved with no prior selection, not all galaxies are expected to be spiral, and not all spiral galaxies provide clear details that can allow to identify the curve of the arms. To remove galaxies that their spin direction cannot be identified, only galaxies that have at least 30 identified peaks aligned in lines are considered as galaxies with identifiable spin directions. The avoid edge-on galaxies, only galaxies that have at least 75\% of their peaks shift in the same directions are used. As discussed in detail in \citep{shamir2011ganalyzer}, Ganalyzer can identify edge-on and elliptical galaxies efficiently. More importantly, it is explainable, and it is not based on any form of machine learning. Detailed information about the annotation can be found in \citep{shamir2011ganalyzer,shamir2012handedness,hoehn2014characteristics,dojcsak2014quantitative,shamir2021particles,shamir2020patterns,shamir2020pasa}.   To ensure that the process of annotation is not affected by differences between different computer systems, all galaxy images were analysed by the exact same code, but also by the exact same computer and exact same processor. The validation that the analysis is not exposed to a possible source of consistent bias is discussed in Section~\ref{error}. 


Some galaxies are part of merging systems, have satellite galaxies, or other large extended objects that overlap or are very close to the galaxy. To remove such objects, objects that had another object within less than 0.01$^o$ were removed. After removing neighbouring objects, the dataset contained 1,287,094 galaxies. Manual inspection of 200 random galaxies showed that no galaxy that was annotated as spinning clockwise seemed by manual inspection to be spinning counterclockwise, and vice versa. The process of annotation was done with the original galaxy images, and then repeated with the mirrored galaxy images by using the {]it flip} command of the {\it ImageMagick} impacge processing tool.

\begin{figure}
\centering
\includegraphics[scale=0.25]{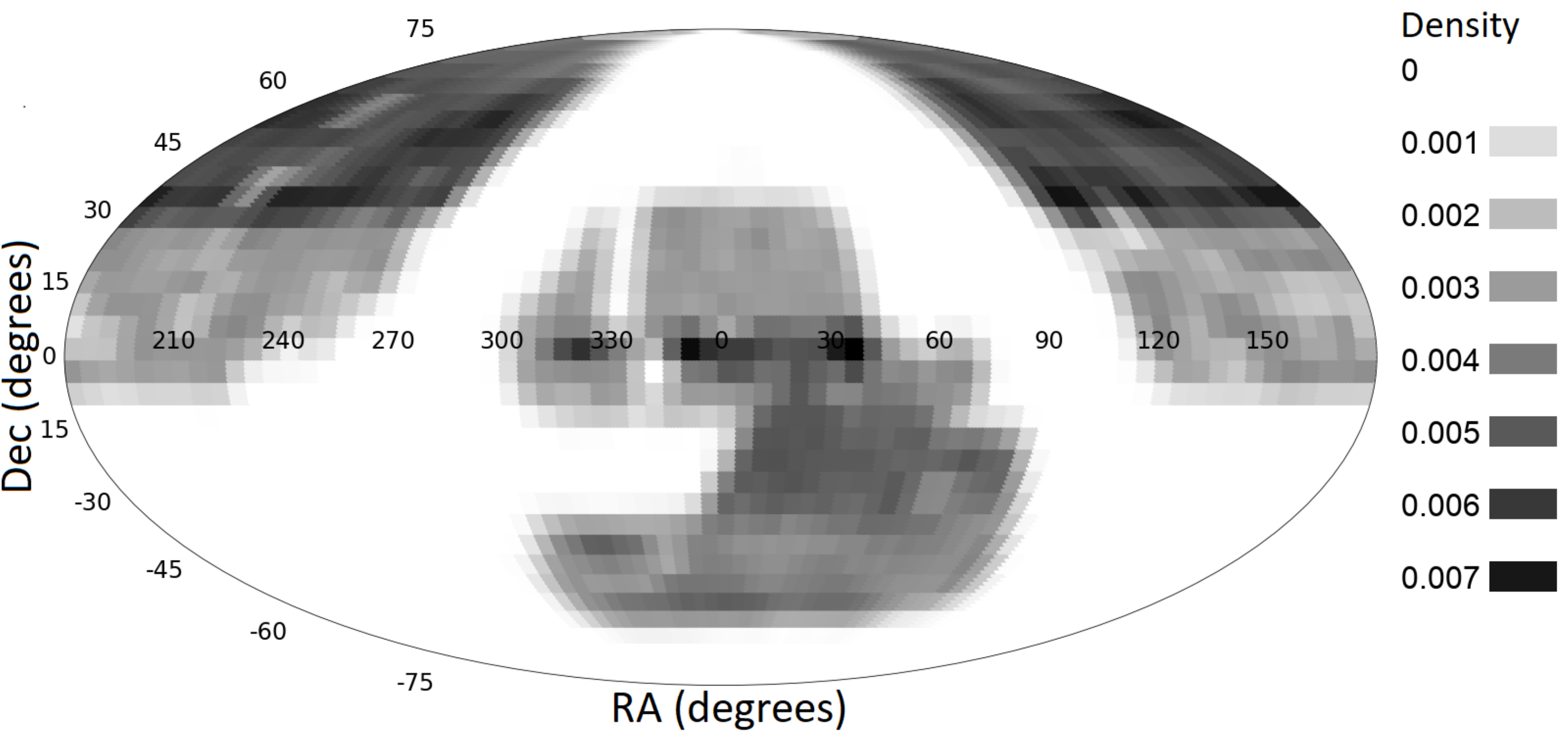}
\caption{The density of the distribution of the galaxy population. The density of the galaxy distribution in each $5^o\times5^o$ field in the sky is determined by the number of galaxies in that $5^o\times5^o$ field divided by the total number of galaxies in the dataset.}
\label{population5}
\end{figure}

Like many other digital sky surveys, galaxies in the DESI Legacy Survey are not distributed uniformly in the sky, and some parts of the sky are not covered at all, especially regions closer to the Galactic plain. Figure~\ref{population5} shows the density of the distribution of the galaxy population in different parts of the sky. The figure shows the number of galaxies in each $5^o\times5^o$ field of the sky divided by the total number of galaxies in the dataset. As expected, the distribution of the density of galaxy population is not uniform in the sky, and some parts of the sky are not imaged at all such as the Milky Way. However, the DESI Legacy Survey provides coverage with footprint larger than any other existing Earth-based sky survey.

\begin{figure}
\centering
\includegraphics[scale=0.7]{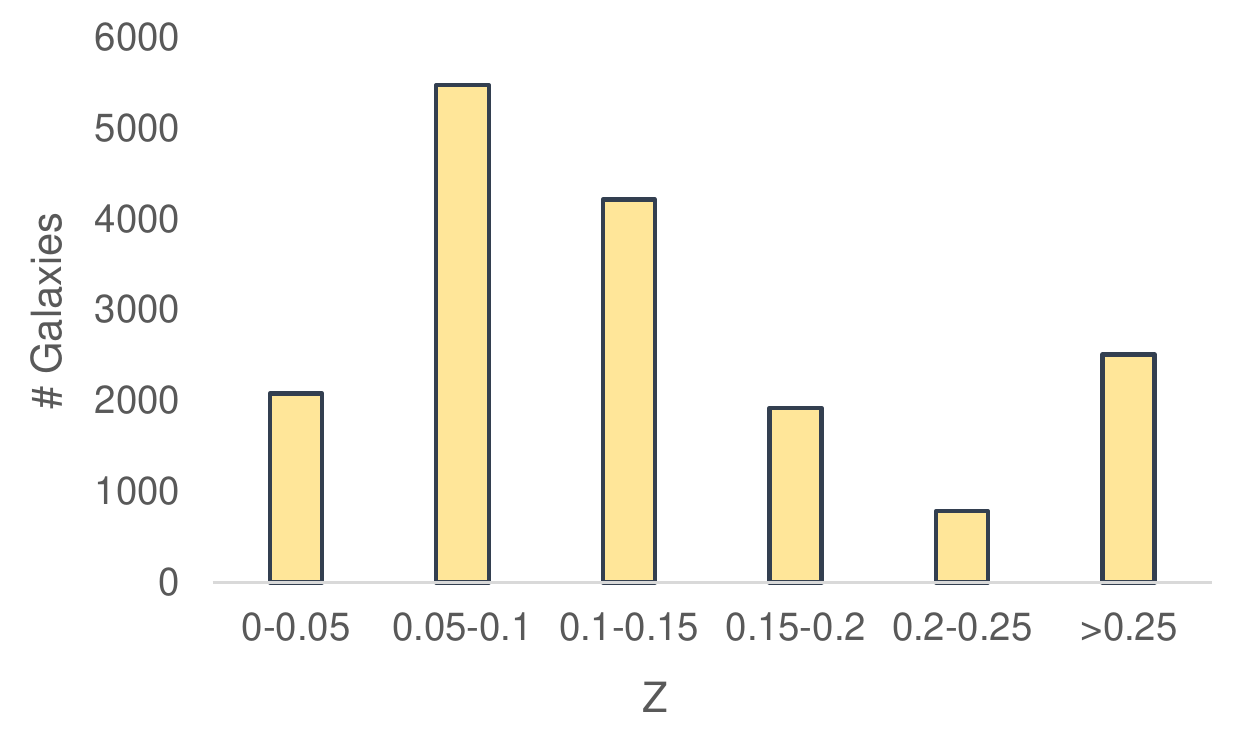}
\caption{The distribution of the redshift of the galaxies in the DESI Legacy Survey. Since DESI Legacy Survey galaxies do not yet have redshifts, the distribution was determined by a subset of galaxies that had spectra through the 2dF redshift survey.}
\label{z_distribution}
\end{figure}

Figure~\ref{z_distribution} shows the distribution of the redshift of the galaxies in the DESI Legacy Survey. The DESI Legacy Survey galaxies do not yet have redshift, and therefore the redshift distribution is determined by a subset of 18,920 galaxies from the DESI Legacy Survey that are also included in 2dF \citep{cole20052df}. 

\section{Results}
\label{results}

Perhaps the simplest way to study the distribution of the spin directions of the galaxies is to separate the galaxies into the Southern and Northern hemispheres, as well as the Eastern and Western hemispheres. Table~\ref{hemispheres2} shows the distribution of the spin directions in several simple separations of the sky. As the table shows, the Northern hemisphere has a higher number of galaxies that spin counterclockwise, while in the Southern hemisphere more galaxies spin clockwise. The excessive number of counterclockwise in the Northern hemisphere is aligned with the previous reports mentioned in Section~\ref{introduction}. Additionally, the table shows an inverse asymmetry in the galaxy spin directions in the Southern hemisphere. Similar observation is also made when the sky is separated into Eastern and Western hemispheres. The asymmetry $A$ was measured by A=$\frac{cw-ccw}{cw+ccw}$.

\begin{table}
\caption{The distribution of galaxy spin directions in different simple opposite hemispheres. The P values are the one-tailed P values of the binomial distribution when assuming random 0.5 probability of a galaxy to spin in either directions.}
\label{hemispheres2}
\scriptsize
\begin{tabular}{lccccc}
\hline
     RA range ($^o$) & Dec range ($^o$) & \# cw          & \# ccw         &  $\frac{\#cw-\#ccw}{\#cw + \#ccw}$  & P  \\ 
\hline
$0^o-360^o$    & -90-0    & 229,900 & 228,461 & 0.00313  & 0.017          \\
$0^o-360^o$    &  0-90 & 411,808 & 416,925 & -0.0062  & $<10^{-5}$ \\
\hline
0-180     & -90-90 & 359,577 & 358,214 & 0.0019 & 0.054 \\
180-360 & -90-90 & 282,131 & 287,172  & -0.0089 & $<10^{-5}$ \\
\hline
\hline
\end{tabular}
\end{table}

Dividing the sky into two hemispheres is a very simple separation, aiming at providing a straightforward statistical analysis. Since the data is not distributed uniformly in the sky, and since the separation of the sky into the Northern and Southern hemisphere is merely an arbitrary separation, that analysis might not be comprehensive enough to provide information regarding the distribution of galaxy spin directions in the sky, if such asymmetry indeed exists.

\begin{figure*}
\centering
\includegraphics[scale=0.6]{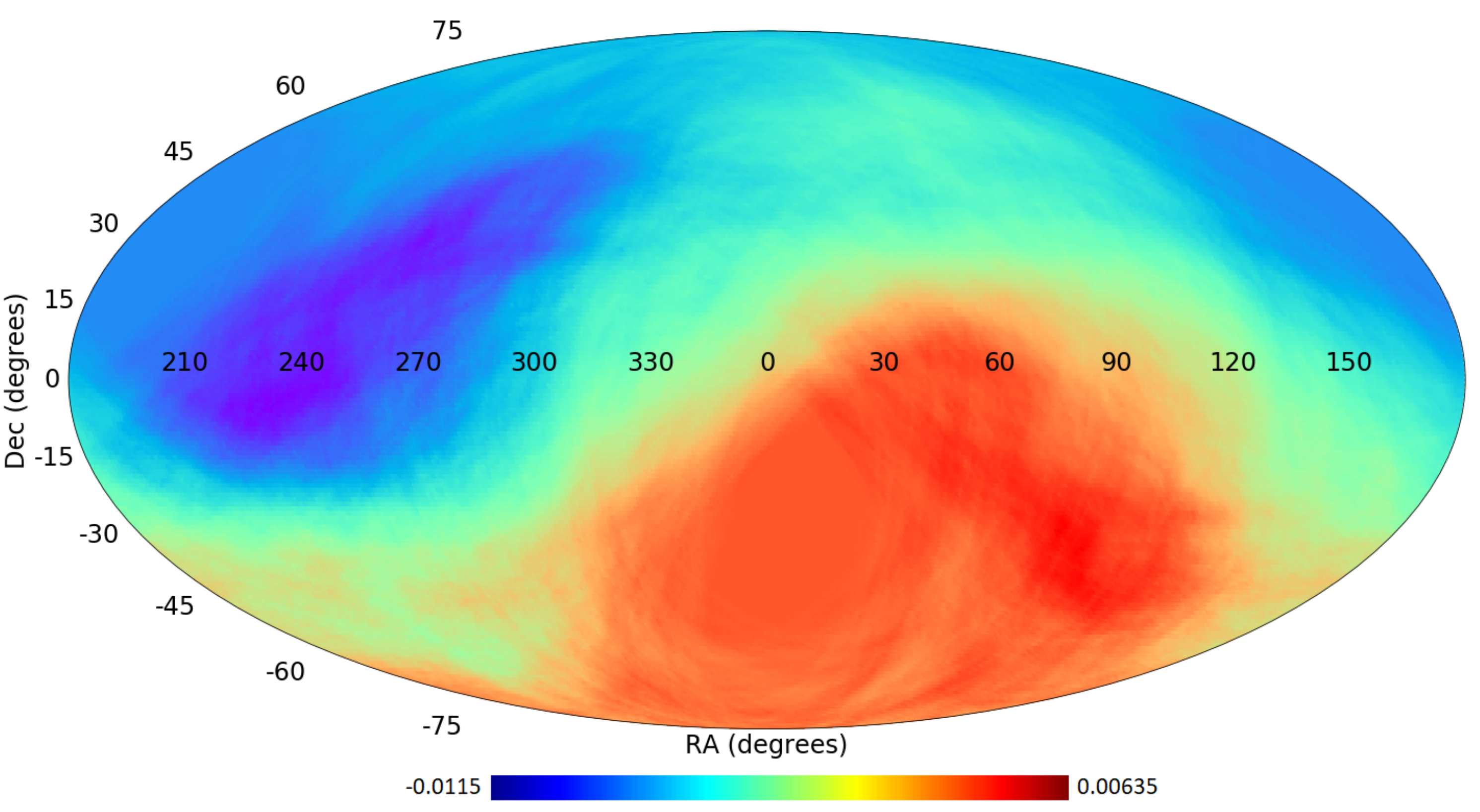}
\caption{The observed asymmetry $\frac{cw-ccw}{cw+ccw}$ in all possible hemispheres. Each $(\alpha,\delta)$ is assigned with the asymmetry between galaxies with opposite spin directions in the hemisphere centred around that location.}
\label{all10_6K_dist90_inc1}
\end{figure*}

To perform an analysis that does not rely on an arbitrary separation to Northern and Southern hemisphere, an analysis was performed such that the ratio between the number of galaxies spinning clockwise and galaxies spinning counterclockwise was measure for each possible hemisphere in the sky. For each integer $(\alpha,\delta)$ combination, the asymmetry $A$ was measured by using all galaxies in the dataset that their angular distance from $(\alpha,\delta)$ is less than 90$^o$. That analysis provides a comparison between all possible separations of the sky into two hemispheres. Figure~\ref{all10_6K_dist90_inc1} shows the results of the analysis centred at each integer $(\alpha,\delta)$ combination in the sky.

The figure shows that the number of spiral galaxies spinning counterclockwise is highest compared to clockwise galaxies at $(\alpha=232^o,\delta=-6^o)$. In the hemisphere centred around $(\alpha=232^o,\delta=-6^o)$ there are 297,099 galaxies spinning clockwise and 304,020 galaxies spinning counterclockwise. Using binomial distribution, the probability to have such distribution or stronger by chance is $P<10^{-17}$ when assuming that a galaxy has a mere chance 0.5 to spin i either directions. The excessive number of galaxies spinning counterclockwise peaks at $(\alpha=84^o,\delta=-29^o$). In that hemisphere, the ratio between clockwise and counterclockwise galaxies is 311,634:307,868. The random probability for that asymmetry or stronger is $P<10^{-6}$. These two peaks are not observed by statistical analysis, but directly from the difference between the number of clockwise and counterclockwise galaxies in all possible hemispheres.

\begin{figure}
\centering
\includegraphics[scale=0.26]{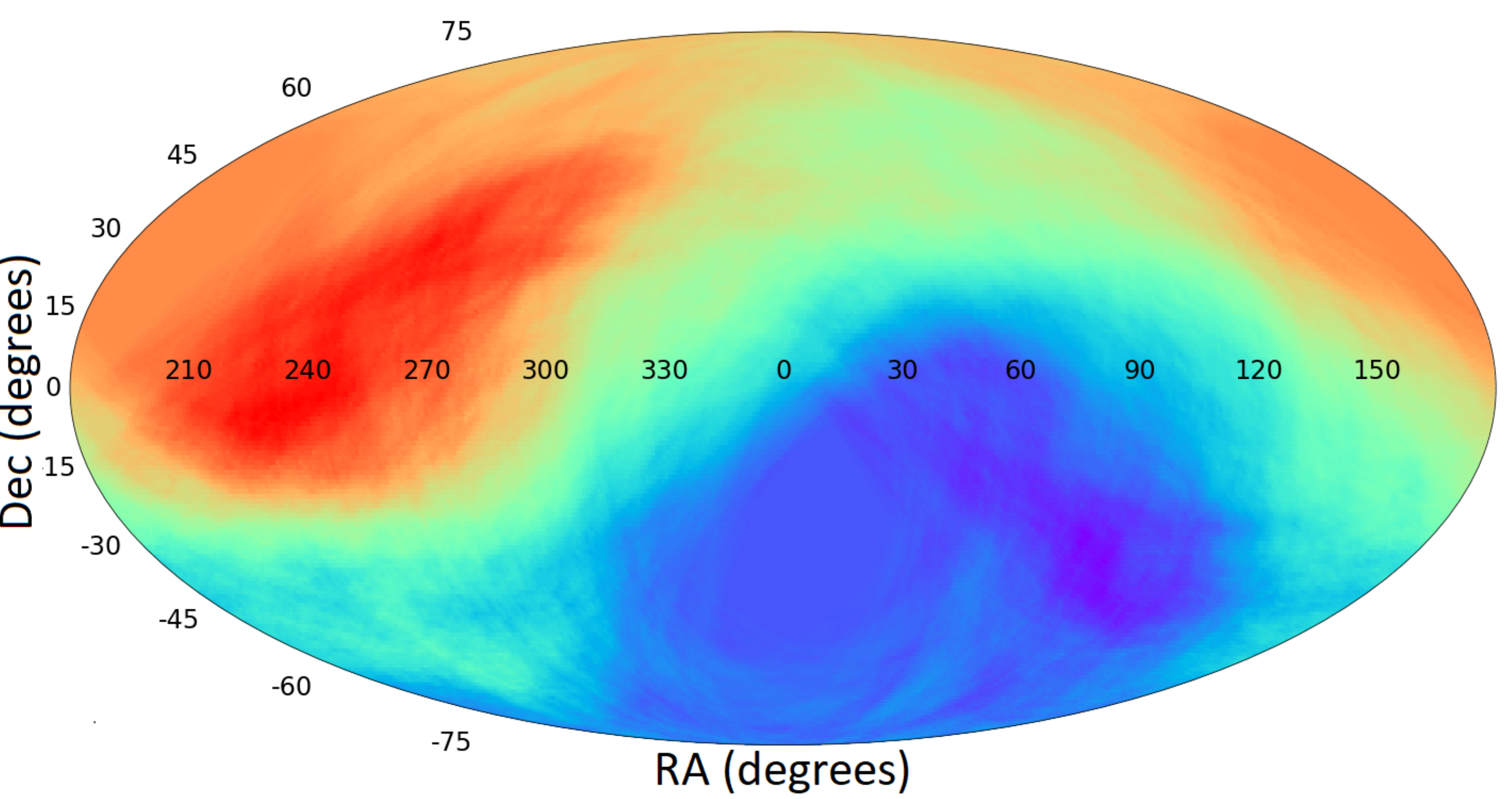}
\caption{The observed asymmetry in all possible hemispheres. when the galaxy images are mirrored.}
\label{all10_6K_dist90_inc1_inverse}
\end{figure}

As a control experiment, the analysis was repeated by mirroring the galaxy images retrieved from the DESI Legacy Survey server, and repeating the entire process with the mirrored images. Images were mirrored using the {\it ImageMagick} image processing tool. Figure~\ref{all10_6K_dist90_inc1_inverse} shows the results after mirroring all images. As expected, the results were exactly inverse compared to the results when using the original images. Since Ganalyzer is fully symmetric, that is expected and aligned with several previous similar experiments \citep{shamir2012handedness,shamir2020patterns,shamir2021large,shamir2022new,shamir2022large,shamir2022possible}. When assigning the galaxies with random spin directions the asymmetry disappears completely, providing a completely flat figure. That is expected since the asymmetry in each $(\alpha,\delta)$ is not the results of a statistical analysis, but just the ratio between the number of galaxies spinning in opposite ways. A collection of random numbers divided by random numbers is always expected to result in a set of random values. 

To apply a statistical analysis, $\chi^2$ statistics was used to identify the location of the strongest dipole axis. That was done by attempting to fit the spin directions of the galaxy to cosine dependence from all possible integer $(\alpha,\delta)$ combinations. 

For the statistical analysis of a possible dipole axis, each galaxy is first assigned with a spin direction $d$, which is {\it 1} if the spin direction of the galaxy spins clockwise, and {\it -1} if the galaxy spins counterclockwise. For each possible $(\alpha,\delta)$ in the sky, the angular distances $\phi+i$ between $(\alpha,\delta)$ and each galaxy {\it i} are computed.
The $\chi_{\alpha,\delta}^2$ from each $(\alpha,\delta)$ integer combination is computed by Equation~\ref{chi2}
\begin{equation}
\chi^2_{(\alpha,\delta)}=\Sigma_i | \frac{(d_i \cdot | \cos(\phi_i)| - \cos(\phi_i))^2}{\cos(\phi_i)} | ,
\label{chi2}
\end{equation}
where $d_i$ is the spin direction (1 or -1) of the galaxy {\it i}, and $\phi_i$ is the angular distance between galaxy {\it i} and $(\alpha,\delta)$.

To compute the statistical significance of the possible axis at $(\alpha,\delta)$, the $\chi^2_{(\alpha,\delta)}$ was also computed 1000 times such that in each run all galaxies were assigned with random spin directions. Using the $\chi^2_{(\alpha,\delta)}$ from 1000 runs, the mean $
\bar{\chi^2}^{random}_{(\alpha,\delta)}$ and standard deviation $\sigma^{random}_{(\alpha,\delta)}$ of the $\chi^2_{(\alpha,\delta)}$ with random spin directions was computed. Then, the statistical signal $\sigma_{(\alpha,\delta)}$ can be determined by Equation~\ref{sigma_difference}

\begin{equation}
\sigma_{(\alpha,\delta)}=\frac{|\chi^2_{\alpha,\delta}-\bar{\chi^2}^{random}_{(\alpha,\delta)}|}{\sigma^{random}_{(\alpha,\delta)}}.
\label{sigma_difference}
\end{equation}

\begin{figure*}
\centering
\includegraphics[scale=0.16]{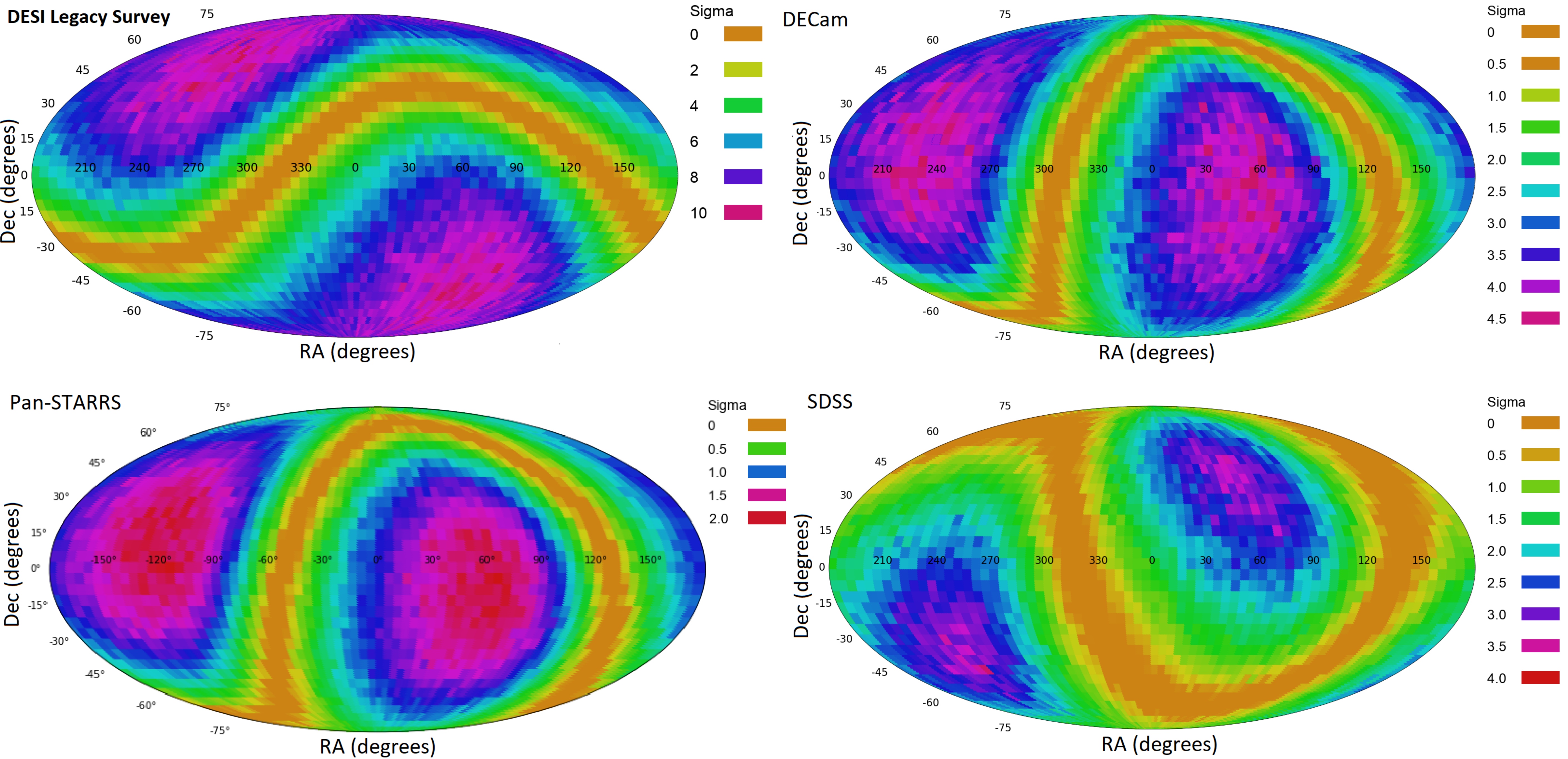}
\caption{The statistical significance of a dipole axis in galaxy spin directions from different $(\alpha,\delta)$. The analysis is compared to previous analyses using DECam \citep{shamir2021large}, Pan-STARRS \citep{shamir2020patterns}, and SDSS \citep{shamir2020patterns}.}
\label{dipole_all}
\end{figure*}

The $\sigma_{(\alpha,\delta)}$ difference between the $\chi^2$ computed with the real spin directions of the galaxies and the mean $\chi^2$ computed with the random spin directions is used to deduce the $\sigma_{(\alpha,\delta)}$ of the $\chi^2$ fitness to occur by chance in each $(\alpha,\delta)$ combination. Figure~\ref{dipole_all} shows the $\sigma_{(\alpha,\delta)}$ of a dipole axis in the data from different $(\alpha,\delta)$ coordinates, as defined by Equation~\ref{sigma_difference}. The most likely axis peaks at $(\alpha=63^o,\delta=-39^o)$, with statistical power of 8.8$\sigma$. The 1$\sigma$ error of the axis is $(-2^o,118^o)$ for the RA, and $(6^o,-90^o)$ for the declination.

The figure also shows a comparison to previous observations with DECam \citep{shamir2021large}, SDSS \citep{shamir2020patterns}, and Pan-STARRS \citep{shamir2020patterns,shamir2022new}. The RA of the most likely axis in all datasets is close, as shown in Table~\ref{axes}, and well within the 1$\sigma$ error range. The declination is somewhat different, although the differences in most cases fall within the 1$\sigma$ error, and therefore the difference between the declination of the most likely axis observed in the different datasets should not be considered statistically significant. Since the declination range in DECam, SDSS, and Pan-STARRS is much smaller than the range of the RA, the declination is expected to have a higher error range, as can also be seen visually in the figure. The dataset that has the strongest difference in its declination from the declination of the axis observed in DESI Legacy Survey is SDSS, with a difference of 1.11$\sigma$. The differences in the declination of the most likely axis in all other datasets are below 1$\sigma$, showing that the dipole axes observed in all datasets are aligned with each other. All observations are statistically significant, except for Pan-STARRS, which is close but below the $2\sigma$ threshold. Pan-STARRS is also by far the smallest dataset, with just 33,028 galaxies. The strongest signal is observed with the DESI Legacy Survey, which is expected due to its large size and its much wider footprint.

\begin{table*}
\centering
\small
\begin{tabular}{lccccc}
\hline
Dataset          &  RA            & Dec          & $\sigma$  & RA 1$\sigma$ error    &  Dec  1$\sigma$ error  \\
                     &  (degrees) & (degrees)  &                & range                        &   range \\
\hline
DESI Legacy Survey  &  63  &  -39   &  8.8  &  -2 - 118      &  6 - -90 \\
DECam              &  57     &  -10   &  4.7  &  22 - 92      &  -39 - 56 \\
SDSS             & 69   & 56   &  4.6   &  19 - 107  &  25 - 77  \\
Pan-STARRS  &   47  &  -1   &  1.9   &  4 - 117  &  -73 - 40  \\
\hline
\end{tabular}
\caption{The most likely dipole axes observed in the different datasets, and the 1$\sigma$ error range of the peak of the axes identified in each dataset.}
\label{axes}
\end{table*}


\begin{figure}
\centering
\includegraphics[scale=0.25]{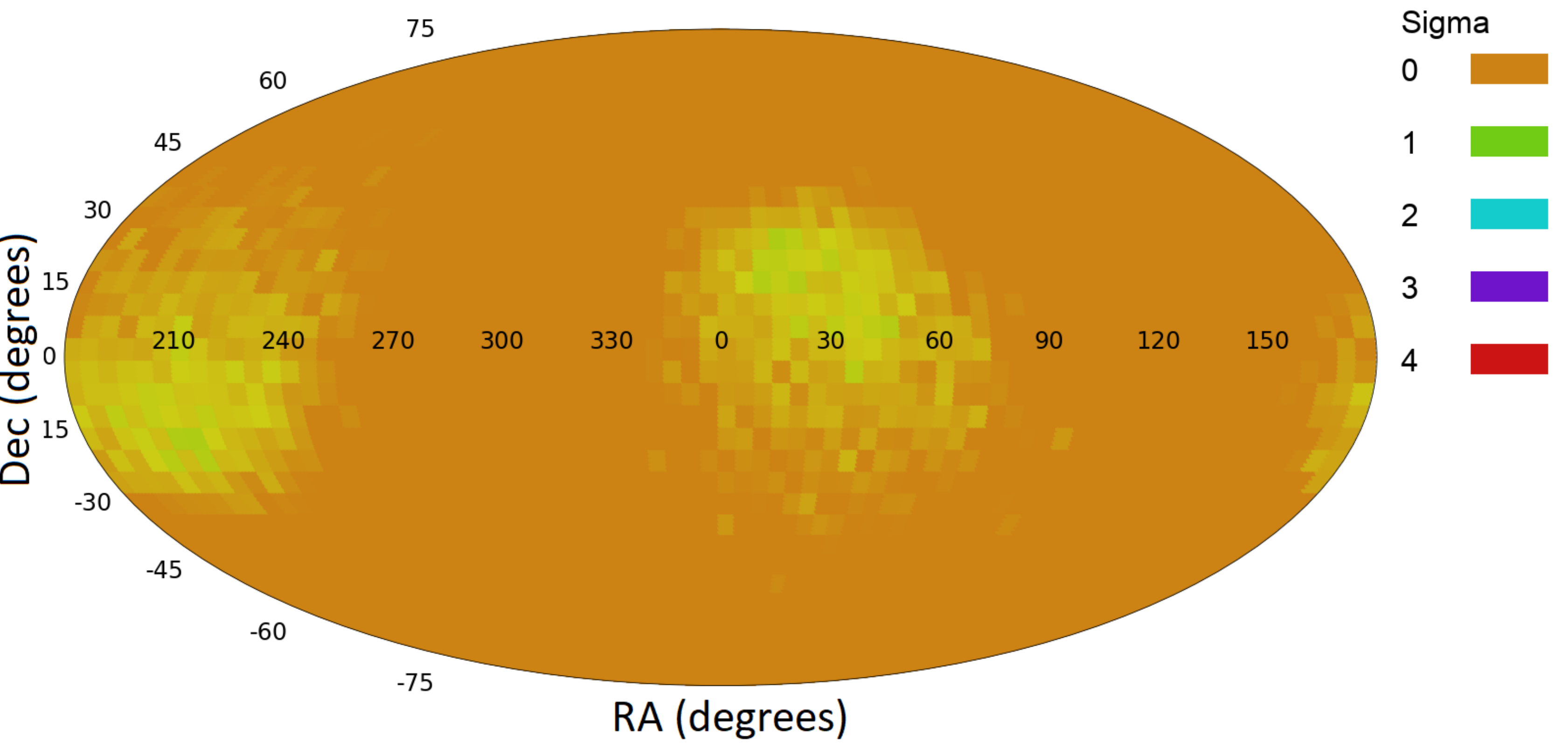}
\caption{The statistical significance of a dipole axis from different $(\alpha,\delta)$ when the spin directions of the galaxies are random.}
\label{dipole_all_random}
\end{figure}

Figure~\ref{dipole_all_random} shows the $\sigma_{(\alpha,\delta)}$ for all integer $(\alpha,\delta)$ combinations when the spin directions of the galaxies are random. The strongest dipole axis in the data has a statistical significance of 0.77$\sigma$. 

\section{Analysis of possible errors}
\label{error}




The purpose of this section is to survey possible sources of error that can explain the observation in a manner that is not of astronomical origin. While such reasons can exist, the purpose of this section is to discuss the known possible causes.

\subsection{Cosmic variance}

Galaxies are not distributed in the Universe in a fully uniform manner. The subtle fluctuations in galaxy population in different parts of the sky can lead to ``cosmic variance'' \citep{driver2010quantifying,moster2011cosmic}. Such variance can affect different measurements at different parts of the sky \citep{kamionkowski1997getting,amarena2018impact,keenan2020biases}. 

The messenger of the difference between galaxies with opposite spin directions is a relative measurement rather than an absolute measurement. That is, the measurement is determined by the differences between two measurements made in the exact same field. The number of galaxies spinning clockwise and the number of galaxies spinning counterclockwise are detected and analysed from the exact same field, exact same instrument, and exact same exposure. Anything that can affect the number of galaxies spinning in one direction will equally affect the number of galaxies spinning in the opposite direction. Unless an unknown or unexpected effect exists, cosmic variance is therefore not expected to affect the measurement.

\subsection{Error in the galaxy annotation algorithm}
\label{symmetry}

A possible source of error is the galaxy annotation algorithm. However, there are several indications that the observation might not be the result of an anomaly in the annotation algorithm. Firstly, the algorithm is a deterministic model-driven algorithm that follows simple defined rules. It is not based on machine learning systems, which can be unexpected and driven by complex biases that these algorithms can learn from the data unexpectedly \citep{dhar2022systematic}.  Clearly, model-driven algorithms can also be biased, and there is a certain possibility that the results shown here are driven by an unknown bias in the algorithm. To reduce that possibility, several experiments were done to test the symmetric nature of the algorithm \citep{shamir2020patterns,shamir2020pasa,shamir2021large,shamir2022new,shamir2022large}. In these experiments the galaxy images were mirrored to ensure that the results are identically inverse to the results observed with the original images.  As also discussed in Section~\ref{results}, an experiment of mirroring the galaxy images used in this study showed identical inverse results, showing empirically that the algorithm is symmetric.

Because the asymmetry is inverse in opposite hemispheres, if the reason was a software error it means that the software error flips in opposite parts of the sky. That makes it unlikely that a software error would exhibit itself in such form. When the images are mirrored, the asymmetry immediately flips.

The main feature of the annotation algorithm for the purpose of this study is its simple ``mechanical'' nature and fully symmetric behavior. Due to the theoretical and empirical analysis that shows that the algorithm is symmetric, an error in the galaxy annotation is expected to impact clockwise and counterclockwise galaxies in a similar manner. If the galaxy annotation algorithm had a certain error in the annotation, the asymmetry {\it A} can be defined by Equation~\ref{asymmetry}.
\begin{equation}
A=\frac{(N_{cw}+E_{cw})-(N_{ccw}+E_{ccw})}{N_{cw}+E_{cw}+N_{ccw}+E_{ccw}},
\label{asymmetry}
\end{equation}
where $E_{cw}$ is the number of galaxies that spin clockwise but were incorrectly annotated as counterclockwise, and $E_{ccw}$ is the number of galaxies that spin counterclockwise but incorrectly annotated as spinning clockwise. Because the algorithm is symmetric, the number of galaxies that spin counterclockwise but were annotated incorrectly as clockwise is expected to be roughly the same as the number of clockwise galaxies missclassified as counterclockwise. Therefore, $E_{cw} \simeq E_{ccw}$ \citep{shamir2021particles}, and the asymmetry {\it A} can be formalized by Equation~\ref{asymmetry2}.

\begin{equation}
A=\frac{N_{cw}-N_{ccw}}{N_{cw}+E_{cw}+N_{ccw}+E_{ccw}}
\label{asymmetry2}
\end{equation}

Because the number of missclassified galaxies cannot be negative, $E_{cw}$ and $E_{ccw}$ must be positive. Therefore, a higher rate of incorrectly annotated galaxies is expected to make the asymmetry {\it A} lower. For that reason, incorrect annotation of galaxies is not expected to lead to asymmetry, and can only make the asymmetry weaker.

An experiment \citep{shamir2021particles} of adding intentional error to the annotation of the galaxies showed that the results do not change substantially even when 25\% of the galaxies are assigned with incorrect spin directions, when the error is added to the counterclockwise and clockwise galaxies \citep{shamir2021particles}. If the error is added only to galaxies that spin clockwise or only to galaxies that spin counterclockwise, even a small error of 2\% leads to a very strong asymmetry of more than $10\sigma$, with a strong dipole axis that peaks exactly at the celestial pole \citep{shamir2021particles}. More information about that experiment can be found in \citep{shamir2021particles}.

Downloading of the image data was done by a single computer to eliminate an unknown possible effect that differences between computer systems can lead to differences in the way the galaxy images are annotated.

\subsection{Bias in the sky survey hardware or photometric pipeline}

Digital sky surveys are some of the more complicated research instruments of our time, involving complex hardware and software to enable the collection, storage, analysis, and accessibility of astronomical data. It is difficult to think of an error in the hardware or software that can lead to asymmetry between the number of clockwise and counterclockwise galaxies, but due to the complexity of these systems it is also difficult to prove that such error does not exist. In this study several different digital sky surveys are analysed independently, all show very similar results. While it is difficult to identify a possible error in one digital sky survey that would exhibit itself in the form of asymmetry shown here, it is difficult to think of such error that is consistent in several different sky surveys. Since such photometric pipelines might share software or algorithms, such possibility, although seems low, should still not be completely ruled out.

\subsection{Multiple photometric objects at the same galaxy}

Digital sky surveys can identify several photometric objects in one galaxy. That can be caused by galaxy mergers, satellite galaxies, large star clusters, detached arm segments, and more. In the datasets all objects that had another object within 0.01$^o$, and therefore all objects that are part of the same galaxy were removed, and a galaxy can only appear once in the dataset. 

It has been shown that even if such objects existed, they are expected to be distributed evenly between clockwise and counterclockwise galaxies, and therefore do not introduce an asymmetry. Experiments with datasets of galaxies assigned with random spin directions showed that adding artificial objects to the galaxies at the exact same position of the original galaxies does not lead to signal of asymmetry \citep{shamir2021particles}. The experiments were done by using $\sim7.7\cdot10^4$ SDSS galaxies assigned with random spin directions. Gradually adding more objects at the exact same locations and spin directions as the original galaxies did not lead to a statistically significant signal \citep{shamir2021particles}.

\subsection{Spiral galaxies with leading arms}

Rarely, the curve of the arms of a spiral galaxy does not provide a reliable information about its spin direction. An example is NGC 4622 \citep{freeman1991simulating}, which has leading galaxy arms. A systematically uneven distribution of such leading arm spiral galaxies might lead to asymmetry in the number of galaxies spinning in opposite directions. For instance, if a high number of galaxies that spin clockwise are backward spiral galaxies, it would lead to a higher number of galaxies annotated as counterclockwise galaxies.

However, spiral galaxies with leading arms that spin in an opposite direction to the arm curves are relatively rare. These galaxies are expected to be distributed evenly among clockwise and counterclockwise galaxies, and there is no indication of more galaxies with leading arm that spin in a certain direction. Therefore, according to the current knowledge, there is no reason to assume that the asymmetry is driven by backward spiral galaxies. The same can also be relevant to multi-spin spiral galaxies \citep{rubin1994multi}. These types of galaxies are also rare, and are also expected to be distributed evenly between galaxies that spin clockwise and galaxies that spin counterclockwise.

\subsection{Atmospheric effect}

There is no known atmospheric effect that can change the spin direction of a galaxy as seen from Earth. Since the asymmetry is determined by galaxies imaged in the same field, any kind of atmospheric effect that affects clockwise galaxies will also affect counterclockwise galaxies in the same way. It is therefore not likely that an atmospheric condition would affect the number of clockwise galaxies imaged in a certain field, but will not affect counterclockwise galaxies. Previous experiments showed the same asymmetry also with the space-based Hubble Space Telescopes \citep{shamir2020pasa}. As a space-based instrument, HST is not supposed to be affected by the Earth atmosphere.

\section{Conclusions}
\label{conclusions}


Multiple studies in the past two decades have shown observational evidence of large-scale alignment of angular momentum of galaxies. The common theory for the origin of galaxy angular momentum is tidal torque theory \citep{hoyle1949problems,peebles1969origin,doroshkevich1970spatial,white1984angular,catelan1996evolution,lee2001galaxy,porciani2002testing,porciani2002testing2,schafer2009galactic,cadiou2021angular}, in which the spin of a galaxy is directly linked to the cosmic initial conditions \citep{cadiou2021angular}. That theory is in agreement with multiple observations of large-scale alignment of spin directions \citep{jones2010fossil,tempel2013evidence,tempel2013galaxy,tempel2014detecting,codis2015spin,pahwa2016alignment,ganeshaiah2018cosmic,ganeshaiah2019cosmic,blue2020chiles,welker2020sami,kraljic2021sdss,lopez2021deviations}. 

The spin directions of spiral galaxies have been shown to be aligned inside cosmic web filaments \citep{tempel2013evidence,tempel2013galaxy,kraljic2021sdss}. Some of these observations are of scales that are far beyond gravitational interactions \citep{lee2019galaxy,lee2019mysterious,cadiou2021angular}. More recent observations showed that cosmic filaments themselves also spin, and the origin of their spin can be explained by angular momenta originating from the Universe initial conditions \citep{sheng2022spin}. 

This paper shows the largest analysis of alignment in galaxy spin directions to date in the sense of the number of galaxies and the size of the footprint. The results show non-random distribution, which is added to previous observations that show large-scale alignment of galaxy spin. That alignment can be viewed as an indication of a link between the galaxy spin and the cosmological initial conditions. The observation reported in the paper might also agree with the contention that the merging of galaxies \citep{vitvitska2002origin,peirani2004angular} is not the sole agent that leads to galaxy angular momentum \citep{cadiou2021angular}. While galaxy mergers can lead to galaxy spin, the results shown here, as well as previous smaller-scale observations, suggest that galaxy mergers or other stochastic gravitational interactions are not the sole reason for galaxy rotation.

Another related observation is the alignment in the polarization of quasars \citep{hutsemekers2014alignment,slagter2022new}. Radio galaxies also showed consistency in their angular momentum \citep{taylor2016alignments}, showing large-scale alignment agreement between surveys such as the Faint Images of the Radio Sky at Twenty-centimetres (FIRST) and the TIFR GMRT Sky Survey (TGSS) \citep{contigiani2017radio,panwar2020alignment}. Large-scale clustering of {\it Fermi} blazars also showed a possible axis alignment \citep{marcha2021large}. In addition to observational studies, numerical simulations also showed links between the large-scale structure and spin directions \citep{hahn2007properties,zhang2009spin,codis2012connecting,trowland2013spinning,libeskind2013velocity,libeskind2014universal,forero2014cosmic,wang2018build,lopez2019deviations}. 

The magnitude of the correlation has been associated with the colour and stellar mass  \citep{wang2018spin,lee2018study}, and that association can also be related to halo formation \citep{wang2017general}, leading to the contention that the spin direction in the halo progenitors is related to the large-scale structure of the early Universe \citep{wang2018build}. 

In addition to cosmological initial conditions, the observation reported in this paper could be related to theories that shift from the standard models. While isotropy is the most common working assumption used in most modern cosmological theories, a growing number of observations using numerous different probes have suggested the existence of large-scale anisotropy \citep{aluri2022observable}. These messengers include short gamma ray bursts \citep{meszaros2019oppositeness}, LX-T scaling \citep{migkas2020probing}, Ia supernova \citep{javanmardi2015probing,lin2016significance}, dark energy \citep{adhav2011kantowski,adhav2011lrs,perivolaropoulos2014large,colin2019evidence}, high-energy cosmic rays \citep{aab2017observation}, quasars \citep{quasars,zhao2021tomographic,secrest2021test}, the distribution of galaxy morphology types \citep{javanmardi2017anisotropy}, or differences in the values of $H_o$ that are not aligned with the standard model \citep{luongo2021larger,dainotti2022evolution}. 

One of the most notable probes that might suggest cosmological-scale anisotropy is the cosmic microwave background radiation \citep{eriksen2004asymmetries,cline2003does,gordon2004low,campanelli2007cosmic,zhe2015quadrupole,ashtekar2021cosmic,yeung2022directional}. The CMB probe is naturally also related to the CMB Cold Spot \citep{cruz655non,mackenzie2017evidence,farhang2021cmb}, which is another statistically significant observation that does not have an immediate explanation according to the standard model. A correlation between higher $H_o$ and the CMB dipole has also been reported \citep{krishnan2021hints,luongo2021larger}.

The accumulating reports on large-scale anisotropy have led to proposed expansions to the standard model, as well as new cosmological theories, or renewed interest in older theories that shift from the standard cosmology. Explanations include primordial anisotropic vacuum pressure \citep{rodrigues2008anisotropic}, moving dark energy \citep{jimenez2007cosmology}, contraction prior to inflation \citep{piao2004suppressing}, double inflation \citep{feng2003double}, multiple vacua \citep{piao2005possible}, or spinor-driven inflation \citep{bohmer2008cmb}. Other theories focused on the geometry of the Universe such as ellipsoidal universe \citep{campanelli2006ellipsoidal,campanelli2007cosmic,campanelli2011cosmic,gruppuso2007complete,cea2014ellipsoidal}. 

An older theory that might have attracted certain new interest is the theory of rotating universe \citep{godel1949example,ozsvath1962finite}. While early rotating Universe models were based on a non-expanding Universe \citep{godel1949example}, these theories were adjusted to explain rotation combined with cosmological expansion  \citep{ozsvath2001approaches,sivaram2012primordial,chechin2016rotation,seshavatharam2020integrated,camp2021}. 

Another older cosmological theory that might fit with the observation of a Hubble-scale axis is black hole cosmology \citep{pathria1972universe}. According to the theory of black hole cosmology, the Universe is the interior of a black hole in another universe \citep{pathria1972universe,stuckey1994observable,easson2001universe,seshavatharam2010physics,poplawski2010radial,tatum2018flat,christillin2014machian,chakrabarty2020toy}. If the Universe is hosted in a black hole, that black hole itself is part of another parent Universe, which is related to the theory of multiverse \citep{carr2008universe,hall2008evidence,antonov2015hidden,garriga2016black,debnath2022anisotropic}, which is one of the older theories in cosmology \citep{trimble2009multiverses,kragh2009contemporary}.

One of the basic observations that show agreement with the black hole cosmological theory is the good agreement between the Schwarzschild radius determined based on the mass of the Universe and the Hubble radius \citep{christillin2014machian}. That agreement can also be considered a coincidence. Another notable observation is the acceleration in the inflation rate of the Universe, which can be explained by black hole cosmology without the need to assume the existence of dark energy. It is interesting that the earlier models of black hole cosmology \citep{pathria1972universe} were proposed before the acceleration in the expansion of the Universe was discovered.

The possible link between the observations made in this paper and black hole cosmology is in the possible existence of a cosmological-scale axis \citep{pathria1972universe,stuckey1994observable,easson2001universe,chakrabarty2020toy}. Black holes spin  \citep{gammie2004black,takahashi2004shapes,volonteri2005distribution,mcclintock2006spin,mudambi2020estimation,reynolds2021observational}, and the spin of black holes is inherited from the star from which they were initially created \citep{mcclintock2006spin}. Therefore, if the Universe is indeed the interior of a black hole in another universe, the Universe is expected to have a preferred axis driven by that initial spin \citep{poplawski2010cosmology,seshavatharam2010physics,seshavatharam2014understanding,christillin2014machian,seshavatharam2020light,seshavatharam2020integrated}. Black hole cosmology is also closely related to theory of holographic universe \citep{susskind1995world,bak2000holographic,bousso2002holographic,myung2005holographic,hu2006interacting,rinaldi2022matrix,sivaram2013holography,shor2021representation}, which can explain the space in our Universe as a projection of the space of the interior of a black hole. Clearly, black hole cosmology or alternative geometrical models of the Universe are at this point unproven theories.

A possible pattern in galaxy spin directions can also be related to the proposed theory of a Universal force field \citep{barghout2019analysis}. The observation that galaxies in opposite lines of sight have opposite spin directions is also aligned with cosmology driven by longitudinal gravitational waves \citep{mol2011gravitodynamics}, according which each galaxy at a certain distance from Earth is expected to have an antipode galaxy under the same physical conditions, but accelerating in the opposite direction \citep{mol2011gravitodynamics}. Other cosmological models proposed that dark energy is anisotropic  \citep{adhav2011kantowski,adhav2011lrs}, and the existence of a cosmological-scale axis can also be related to holographic big bang \citep{pourhasan2014out,altamirano2017cosmological}.

This paper showed an analysis of possible differences between the number of galaxies spinning in opposite directions in different parts of the sky. The probe of asymmetry in galaxy spin directions has an advantage of being a relative measurement, and therefore not expected to be affected by background contamination such as Milky Way obstruction. Spiral galaxies are obviously very common in the Universe, allowing to study the distribution in different parts of the sky, and with a large number of instances that can lead to strong statistical signal. 

The DESI Legacy Survey provides image data for a large number of galaxies and a footprint substantially larger than any previous study of this kind. Its downside is that at this point the Dark Energy Spectroscopic Instrument does not provide spectroscopy data for these galaxies, making it more difficult to make a direct analysis of specific filaments and walls in the large-scale structure. Spectroscopic surveys such as 2MASS \citep{huchra20122mass} are far less powerful than the future DESI. For instance, 2MASS has spectra for just a small portion of only 1,405 of the galaxies in the Southern hemisphere used in this study. The availability of redshift of a large number of galaxies as expected by DESI will allow to expand previous studies \citep{hess2013simulating,courtois2013cosmography,jasche2019physical,mcalpine2022sibelius} by analysis of the large-scale distribution of spin directions with the filaments, clusters, and walls of the large-scale structure of the local Universe.


\section*{Acknowledgments}
              
I would like to thank the knowledgeable anonymous reviewer for the insightful comments. The research was funded in part by NSF grant AST-1903823.

\section*{Data Availability}

The source of the data is the DESI Legacy Survey \url{https://www.legacysurvey.org/}, and SDSS. Annotated SDSS data discussed in this study are available at \url{https://people.cs.ksu.edu/~lshamir/data/assym_72k/}. Annotated DESI Legacy Survey data will be provided upon reasonable request.



\bibliographystyle{apalike}

\bibliography{dipole_des} 




\bsp	
\label{lastpage}
\end{document}